\documentclass[aps,prl,twocolumn,amsart,amsmath,amssymb,floatfix,longbibliography]{revtex4-2}

\usepackage[T1]{fontenc}
\usepackage{blindtext}
\usepackage{lineno}
\usepackage{graphicx}
\usepackage{dcolumn}
\usepackage{bm}
\usepackage[normalem]{ulem} 
\usepackage{amsmath,bm}
\usepackage{amsfonts}
\usepackage{amssymb,enumerate}
\usepackage{graphicx,hyperref}
\usepackage{amssymb}
\usepackage{epstopdf,booktabs}
\usepackage{empheq}
\usepackage{fancybox}
\usepackage{ulem}
\usepackage{amssymb}
\usepackage[usenames,dvipsnames]{color}
\usepackage[utf8]{inputenc}
\usepackage{natbib}
\usepackage{soul}
\hypersetup{hidelinks,colorlinks=true,citecolor=blue,linkcolor=blue,urlcolor=black}

\newcommand*{\rom}[1]{\expandafter\@slowromancap\romannumeral #1@}

\begin{document}
\title{Phase-Matched Harmonic Generation in Strongly Magnetized Plasma}

\author{Sida Cao}
\email{sidacao@stanford.edu}
\affiliation{Department of Mechanical Engineering, Stanford University, Stanford, California 94305, USA}
\author{Matthew R. Edwards}
\email{mredwards@stanford.edu}
\affiliation{Department of Mechanical Engineering, Stanford University, Stanford, California 94305, USA}

\date{\today}

\begin{abstract}
Harmonic generation in underdense spatially homogeneous plasma is generally expected to be inefficient: in an unmagnetized uniform plasma the fundamental and its harmonics cannot be phase-matched, resulting in third-harmonic generation efficiencies of no more than $10^{-5}$.
Here, we describe how a strong uniform magnetic field allows phase-matched harmonic generation in constant-density plasma.
We derive phase-matching relations for Type I and Type II second-, third-, and fourth-harmonic generation, and confirm these relations with particle-in-cell simulations.
These simulations show that for weakly relativistic femtosecond pulses the efficiencies of second-, third-, and fourth-harmonic generation can reach at least 70\%, 14\%, and 2\% respectively.                       
Additionally, if driven by a two-color beam, third harmonic generation is found to be over 70\% efficient and fourth harmonic generation is found to be over 30\% efficient. 
\end{abstract}

\maketitle

\section*{Introduction}
Laser pulses propagating through plasma at near-relativistic intensity can drive electron oscillations with significant nonlinearity, generating odd harmonics of the fundamental frequency \cite{sprangle1990nonlinear,mori1993relativistic,esarey1993nonlinear,gibbon1997high}.
The phase velocity mismatch between the fundamental and its harmonics limits the overall efficiency of harmonic generation~\cite{rax1992third}, just as it does for harmonic generation in gaseous media; third-harmonic conversion efficiency in underdense plasma is less than $\sim 10^{-5}$~\cite{mori1993relativistic}.
Harmonic generation in underdense plasmas cannot be phase-matched because the phase velocity of an ordinary electromagnetic wave in a plasma monotonically decreases with increasing frequency~\cite{shkolnikov1993phase}.
Harmonic and high-order harmonic generation in plasmas was explored for frequency conversion of high-power laser beams, as plasmas support far higher intensities than gases \cite{krausz2009attosecond}, but the inefficiency of harmonic generation in both magnetized and unmagnetized underdense plasma has precluded most applications. 
Although only odd harmonics can be generated in symmetric media, the addition of a magnetic field to a plasma breaks the inversion symmetry and allows the generation of even harmonics in addition to odd harmonics, depending on the magnetic field direction and the laser polarization~\cite{dhalia2023harmonic,maity2021harmonic}. 
The presence of a strong magnetic field also enhances the nonlinear response of a plasma \cite{jha2007second}.
Although magnetization increases the harmonic generation efficiency~\cite{jha2006self,jha2007second,kant2016second,kant2016second,avinash2004theory}, phase mismatch remains a severe constraint, limiting the efficiency of second and third harmonic generation to around $10^{-3}$.

In an effort to overcome the efficiency saturation imposed by phase mismatch, several quasi-phase-matching schemes have been proposed, including rippling the plasma density~\cite{rax1992third,rax1993phase,devi2018resonant,dahiya2007phase,liu2008third,kuo2007enhancement} and
spatially varying a magnetic~\cite{rax2000relativistic,abedi2023relativistic,abedi2018second} or electric~\cite{verma2015phase} field.
Phase-matched difference-frequency generation has also been demonstrated using two-color beams to produce ninth harmonic light~\cite{meyer1996phase}, and frequency-dependent light propagation in a gas fiber waveguide was utilized to achieve phase-matched generation of x-rays~\cite{rundquist1998phase}.
However, a mechanism for perfect phase-matching of optical harmonic generation in a uniform plasma has not previously been described. 

\begin{figure}[t]
    \centering
    \includegraphics[width=1.0\linewidth]{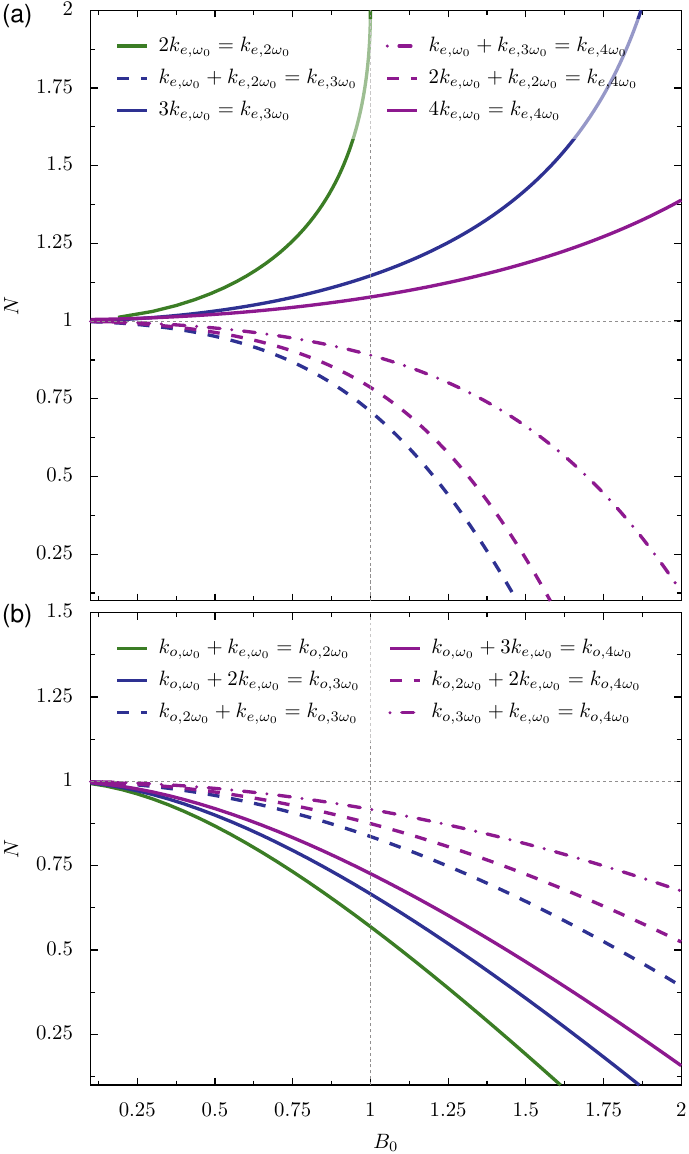}
    \caption{Phase matching conditions for (a) Type I harmonic generation processes. (b) Type II harmonic generation processes.}
    \label{fig:PMconditions}
\end{figure}

In this work, we show that phase-matched harmonic generation can be achieved in a sufficiently magnetized, uniform plasma by utilizing the ordinary mode (O-mode) and the two branches of the extraordinary mode (X-mode).
These two the two electromagnetic modes in magnetized plasmas if the magnetic field is transverse to the propagation direction~\cite{nicholson1983introduction}. 
%
%
When the cyclotron frequency approaches the laser frequency, the O-mode and both branches of the X-mode become closer on the dispersion diagram, as shown in Fig.~\ref{fig:OEdispersion}, enabling the fundamental and its harmonics to appear as either distinct modes or on separate branches of the same mode.
The different dispersion relations for these modes can be tuned by varying the magnetic field strength and the plasma density to allow phase-matching in a manner directly analogous to Type I and Type II harmonic generation in crystals. 
Type I phase-matching involves only extraordinary photons and Type II involves both ordinary and extraordinary photons.
As shown in Fig.~\ref{fig:PMconditions}, this produces two distinct phase-matching conditions for second harmonic generation (SHG). We also find analytic conditions for third (THG) and fourth harmonic generation (FHG), where there are four and six distinct mechanisms, respectively.

\section{Harmonic Generation in Magnetized Plasma}
We consider a light wave in a strongly magnetized $(\sim0.1\ \mathrm{GG})$ uniform plasma, where the light propagates in the $z$ direction, its polarization is in the $y$ direction, and the magnetic field is in the $x$ direction. 
Following~\cite{jha2007second}, we linearize the fluid equations and Maxwell's equations to find the first-order quantities and the nonlinearity that contribute to second harmonic generation.
The fluid equations and Maxwell's equations are:
\begin{subequations}
\label{eq:fluid}
\begin{eqnarray}
        \frac{\partial n}{\partial t} + \nabla\cdot(n\mathbf{v}) &=& 0 \\
        \frac{\partial \mathbf{v}}{\partial t} + (\mathbf{v}\cdot\nabla)\mathbf{v} &=& -\frac{e}{m_e}\left(\mathbf{E} + \mathbf{v}\times \mathbf{B}\right) \\
        \nabla^2\mathbf{E} - \frac{1}{c^2}\frac{\partial^2 \mathbf{E}}{\partial t^2} &=& \mu_0\frac{\partial \mathbf{J}}{\partial t}
\end{eqnarray}
\end{subequations}
where $n$ is the plasma electron density, $\mathbf{v}$ is the electron velocity, $t$ is time, $e$ is the elementary charge, $m_e$ is electron mass, $\mathbf{E}$ is the electric field, $\mathbf{B}$ is the magnetic field, $c$ is the speed of light, $\mu_0$ is the vacuum permeability, and $\mathbf{J} = - en\mathbf{v}$ is the total current assuming immobile ions. 
Assuming small perturbations, we expand the quantities as $n = n_{0} + n_{1}$, $\mathbf{v} = \mathbf{v_1}$, $\mathbf{E} = \mathbf{E}_1$, and $\mathbf{B} = \mathbf{\Tilde{B}_0} + \mathbf{\Tilde{B}_1}$, where the subscript $0$ and $1$ indicate averaged quantities and first-order small perturbations, $n_{0}$ is the averaged plasma density, and $\mathbf{\Tilde{B}_0} = \Tilde{B}_0\hat{x}$ is the applied magnetic field.
The light wave takes the form $\mathbf{E_1} = E_1\cos{(k_0z-\omega_0 t)}\hat{y}$, where $E_1$ is the amplitude of the electric field, $k_0 = 2\pi/\lambda_0$ is the vacuum wavenumber, $\lambda_0$ is the wavelength, and $\omega_0$ is the frequency.
%
%
%
%
%
%
%
\begin{figure}[t]
    \centering
    \includegraphics[width=1\linewidth]{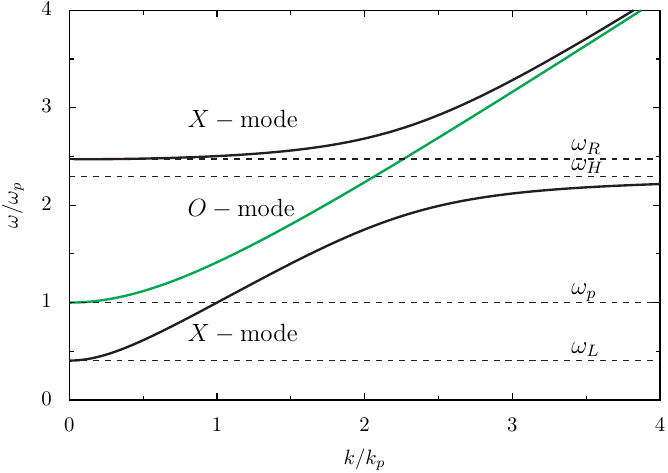}
    \caption{Dispersion relations for the extraordinary mode (X-mode) and the ordinary mode (O-mode) with $N = 0.6$ and $B_0 = 1.6$.}
    \label{fig:OEdispersion}
\end{figure}
Linearizing Eq.~\ref{eq:fluid} gives:
\begin{subequations}
\label{eq:first_order}
\begin{eqnarray}
        v_{z1} &=& -\frac{a_0c\omega_c\omega_0}{\omega_0^2 - \omega_H^2}\cos{(k_0z - \omega_0t)}\\[8pt]
        v_{y1} &=& \frac{a_0c(\omega_0^2-\omega_p^2)}{\omega_0^2 - \omega_H^2}\sin{(k_0z - \omega_0t)}\\[8pt]
        n_{1} &=& -\frac{n_{e0}a_0k_0c\omega_c}{\omega_0^2 - \omega_H^2}\cos{(k_0z - \omega_0t)}
\end{eqnarray}
\end{subequations}
where $a_0 = eE_1/m_e\omega_0c$ is the normalized vector potential, $\omega_p^2 = n_{0}e^2/\epsilon_0m_e$ is the plasma frequency, $\epsilon_0$ is the vacuum permittivity, $\omega_c = e\Tilde{B_0}/m_e$ is the cyclotron frequency, and $\omega_H = (\omega_c^2 + \omega_p^2)^{1/2}$ is the upper hybrid frequency. 
The first order nonlinear current at the fundamental frequency is $J_{y1} = -en_{0}v_{y1}$.

Plugging the first order current into Eq.~\ref{eq:fluid}c, we find the well-known dispersion relation of the extraordinary mode:
\begin{equation}
    \label{eq:x_wave_dispersion}
    n_{e}^2 = \frac{c^2k^2}{\omega^2} = 1 - \frac{\omega_p^2}{\omega^2}\frac{\omega^2-\omega_p^2}{\omega^2 - \omega_H^2}
\end{equation}
where $n_{e}$ is the refractive index for the extraordinary mode and $k$ is the wavenumber. Following the same method, we can derive the dispersion relation for the ordinary mode, where the magnetic field is aligned with the laser polarization~\cite{nicholson1983introduction}:
\begin{equation}
    \label{eq:o_dispersion_relation}
    n_{o}^2 = \frac{c^2k^2}{\omega^2} = 1 - \frac{\omega_p^2}{\omega^2}
\end{equation}
where $n_o$ is the refractive index for the ordinary mode. As shown in Fig.~\ref{fig:OEdispersion}, there are two X-mode branches and one O-mode branch, where $\omega_L = 1/2\left[-\omega_c + (\omega_c^2+4\omega_p^2)^{1/2}\right]$ and $\omega_R = 1/2\left[\omega_c + (\omega_c^2+4\omega_p^2)^{1/2}\right]$ are the left-hand and right-hand cutoff frequencies, respectively. 
%


\begin{figure}[t]
    \centering
    \includegraphics[width=1.0\linewidth]{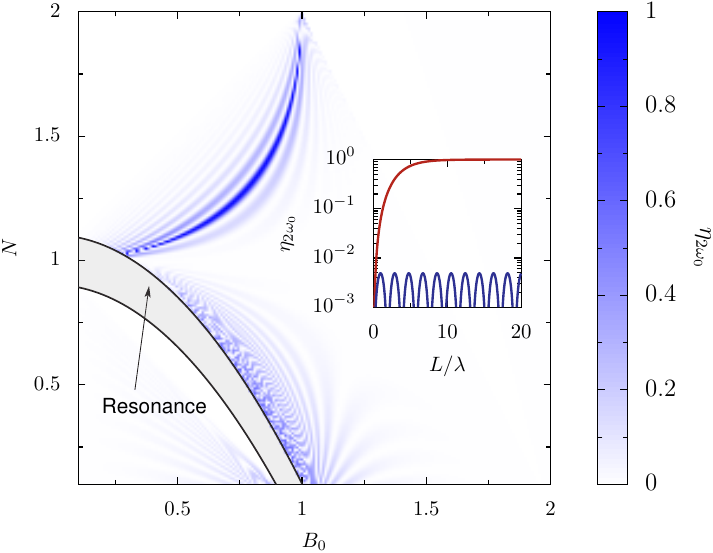}
    \caption{Analytic calculation of Type I second harmonic conversion efficiency with a propagation distance $L = 10\lambda_0$, treating the magnetized plasma as a crystal with a nonlinear susceptibility $\chi^{(2)}_{\mathrm{I}}$ given in Eq.~\ref{eq:chi2_definition}. The gray area is removed due to resonance. The inset shows the conversion efficiency at varied propagation distances. $N = 1.31$ and $B_0 = 0.8$ for the phase-matched case shown in red. $N = 0.75$ and $B_0 = 1.2$ for the phase-mismatched case shown in blue. }
    \label{fig:eff_second_analytic}
\end{figure}

%
We next examine why second harmonic generation happens with the first order quantities given in Eq.~\ref{eq:first_order}.
From Eq.~\ref{eq:fluid}c, we see that a current $\mathbf{J}$ oscillating at frequency $2\omega_0$ is essential for second harmonic generation.
%
%
The nonlinear current in the $y$ direction is $J_{y,2\omega_0}^{NL} = -en_{1}v_{y1}$, which is a result of the coupling between the density oscillation at $\omega_0$ and the transverse electron quiver motion at $\omega_0$~\cite{jha2007second}, where the subscript $2\omega_0$ means oscillation at $2\omega_0$.
This coupling disappears if the magnetic field is zero, since the density perturbation is proportional to the cyclotron frequency $n_{1}\propto \omega_c$, as can be seen from Eq.~\ref{eq:first_order}c.
%
%
%

To understand how this current drives second harmonic generation, we treat the magnetized plasma as a lossless medium with a nonlinear susceptibility $\chi^{(2)}_{\mathrm{I}}$.
Here we consider Type I processes first.
%
%
%
We rewrite the wave equation Eq.~\ref{eq:fluid}c with the free charge current replaced by a polarization field:
\begin{equation}
\label{eq:forced_wave_equation_with_polarization}
    \nabla^2 \mathbf{E} - \frac{1}{c^2}\frac{\partial^2\mathbf{E}}{\partial t^2} = \mu_0\frac{\partial^2\mathbf{P} }{\partial t^2}
\end{equation}
where $\mathbf{P}$ is the polarization field.
Compared to Eq.~\ref{eq:fluid}c, the time derivative of the polarization field plays the same role as the free charge current.
We can determine a polarization field that has an equivalent effect as the current:
\begin{equation}
    \label{eq:nonlinear_polarization}
    \begin{split}
        P_{y,2\omega_0} &= \int J_{y,2\omega_0}dt = -e \int\left( n_{0}v_{y,2\omega_0} +  n_{1}v_{y1}\right)dt\\
    & = P_{y,2\omega_0}^{L} + P^{NL}_{y,2\omega_0}
    \end{split}
\end{equation}
where $v_{y,2\omega_0}$ is the velocity perturbation at twice the fundamental frequency, which can be derived by substituting $\omega_0\to2\omega_0$, $k_0\to k_{2\omega_0}$ in the expression for $v_{y1}$, and $k_{2\omega_0}$ is the wavenumber of the second harmonic. 
$P_{y,2\omega_0}^L = \epsilon_0\chi^{(1)}_{\mathrm{I}}E_1$ and $P^{NL}_{y,2\omega_0} = \epsilon_0\chi^{(2)}_{\mathrm{I}}E_1^2$ are the linear and nonlinear polarization fields at frequency $2\omega_0$, $\chi^{(1)}_{\mathrm{I}}$ is the linear susceptibility, and $\chi^{(2)}_{\mathrm{I}}$ is the nonlinear susceptibility.
$\chi^{(1)}_{\mathrm{I}}$ is directly related to the refractive index given in Eq.~\ref{eq:x_wave_dispersion}, where $\chi^{(1)}_{\mathrm{I}} = n_{\mathrm{e},\omega_0}^2 - 1$, and $n_{e,\omega_0}$ is the refractive index of the extraordinary mode at frequency $\omega_0$.

We move $\chi^{(1)}_{\mathrm{I}}$ to the left and leave only the nonlinear polarization field on the right as the source.
The $\hat{y}$ component of Eq.~\ref{eq:forced_wave_equation_with_polarization} can then be rewritten as:
\begin{equation}
\label{eq:forced_wave_equation_with_polarization_nonlinear}
    \nabla^2 E - \frac{\epsilon_r}{c^2}\frac{\partial^2E}{\partial t^2} =\mu_0\frac{\partial^2P_{y,2\omega_0}^{NL} }{\partial t^2}
\end{equation}
where $\epsilon_r = 1 + \chi^{(1)}_{\mathrm{I}}$ is the relative permittivity.
%
%
The nonlinear polarization field $P^{NL}_{y,2\omega_0} = -e\int n_{1}v_{y1}dt$ can be determined with the first-order quantities given in Eq.~\ref{eq:first_order}.
By definition, the nonlinear susceptibility $\chi^{(2)}_{\mathrm{I}}$ is then:
\begin{equation}
    \label{eq:chi2_definition}
    \chi^{(2)}_{\mathrm{I}} = \frac{2P^{NL}_{y,2\omega_0}}{\epsilon_0E_1^2} = -\frac{2n_{e0}e^3}{\epsilon_0m_e^2}\frac{k_0\omega_c(\omega_0^2 - \omega_p^2)}{\omega_0^3(\omega_0^2 - \omega_H^2)^2}
\end{equation}
where an additional factor of 2 in Eq.~\ref{eq:chi2_definition} comes from the fact that here we are considering second harmonic generation with the same input polarizations, so there is only one indistinguishable permutation~\cite{boyd2008nonlinear}. 
In normalized units, $\chi^{(2)}_{\mathrm{I}}$ can be written as:
\begin{equation}
    \label{eq:chi2_definition_normalized}
    \chi^{(2)}_{\mathrm{I}}\left[\frac{m_e\omega_0c}{e}\right] = -2n_{e,\omega_0}\frac{N(1 - N)B_0}{(1 - N - B_0^2)^2}
\end{equation}
where $N = n_{0}/n_c$ is the normalized plasma density, $n_c = \epsilon_0 m_e\omega_0^2/e^2$ is the critical density, and $B_0 = e\Tilde{B}_0/m_e\omega_0$ is the normalized magnetic field strength.
Following the same method, the nonlinear susceptibility for Type II second harmonic generation $\chi^{(2)}_{\mathrm{II}}$ is:
\begin{equation}
    \label{eq:chi2_definition_normalized_typeII}
    \chi^{(2)}_{\mathrm{II}}\left[\frac{m_e\omega_0c}{e}\right] = -n_{e,\omega_0}\frac{NB_0}{1 - N - B_0^2}
\end{equation}
%

With this plasma-crystal analogy, we can gain some insight into the conversion behavior from our understanding of second harmonic generation in crystals.
Second harmonic generation in crystals is usually described using the coupled-wave equation under the slowly varying envelope assumption~\cite{boyd2008nonlinear}:
\begin{subequations}
\label{eq:coupled_wave_equation}
\begin{eqnarray}
        \frac{d\mathcal{E}_{2\omega_0}}{dz} &=& -i\frac{\omega_0\chi^{(2)}}{2cn_{e,2\omega_0}}\mathcal{E}_{\omega_0}^2e^{i\Delta k z}
        \\[8pt]
        \frac{d\mathcal{E}_{\omega_0}}{dz} &=& -i\frac{\omega_0\chi^{(2)}}{2cn_{e,\omega_0}}\mathcal{E}_{2\omega_0}\mathcal{E}_{\omega_0}^*e^{-i\Delta kz}
\end{eqnarray}
\end{subequations}
where $\mathcal{E}_{\omega_0}$ and $\mathcal{E}_{2\omega_0}$ are the envelopes of the fundamental and the second harmonic respectively, $n_{e,2\omega_0}$ is the refractive index of the extraordinary mode at frequency $2\omega_0$, $\Delta k = k_{e,2\omega_0} - 2k_{e,\omega_0}$ is the phase-mismatch, and $k_{e,\omega_0}$ and $k_{e,2\omega_0}$ are the wavenumbers of the extraordinary mode at frequency $\omega_0$ and $2\omega_0$. 
%
%
Second harmonic generation is expected to be more efficient with minimum phase-mismatch.
%

%
Figure~\ref{fig:eff_second_analytic} shows an analytic calculation of Type I second harmonic conversion efficiency $\eta_{2\omega_0}$ at varied plasma densities and magnetic field strengths.
Note that the region around the upper hybrid resonance is removed, because the lossless assumption breaks and absorption has to be considered.
We see that there exists a well-defined efficiency peak.
We pick two cases to examine the growth of the second harmonic, as shown in the inset of Fig.~\ref{fig:eff_second_analytic}.
%
%
In the phase-mismatched case where $N = 0.75$ and $B_0 = 1.2$, the conversion efficiency shows oscillatory behavior and the maximum conversion efficiency is on the order of $10^{-3}$.
In the phase-matched case where $N = 1.31$ and $B_0 = 0.8$, the efficiency monotonically increases with the propagation distance and is three orders of magnitude higher than the phase-mismatched case.
%
%
%
The coherence length, which is half of the oscillation period, is defined as $l_c = \pi/\Delta k$.
%
With $N = 0.75$ and $B_0 = 1.2$, the phase-mismatch is $\Delta k \approx 0.52k_0$.
The coherence length is $l_c \approx 0.96\lambda_0$, which matches the oscillation period shown with the blue line in the inset of Fig.~\ref{fig:eff_second_analytic}.
With $N = 1.31$ and $B_0 = 0.8$, the phase-mismatch is $\Delta k \approx 0.003k_0$, which is two orders of magnitude smaller.

\section{Phase-matched second harmonic generation}
In this section, we examine the phase-matching conditions for second harmonic generation in magnetized plasma.
Two types of phase-matching processes are presented:
Type I phase-matching converts two X-mode photons on the lower branch at frequency $\omega_0$ to one X-mode photon on the upper branch at frequency $2\omega_0$.
The phase-matching condition requires $N > 1$ and $B_0 < 1$.
Type II phase-matching converts one X-mode photon on the lower branch and one O-mode photon both at frequency $\omega_0$, to one O-mode photon at frequency $2\omega_0$.
Unlike Type I, Type II phase-matching requires underdense plasmas.
%
%
\begin{figure}[t]
    \centering
    \includegraphics[width=1\linewidth]{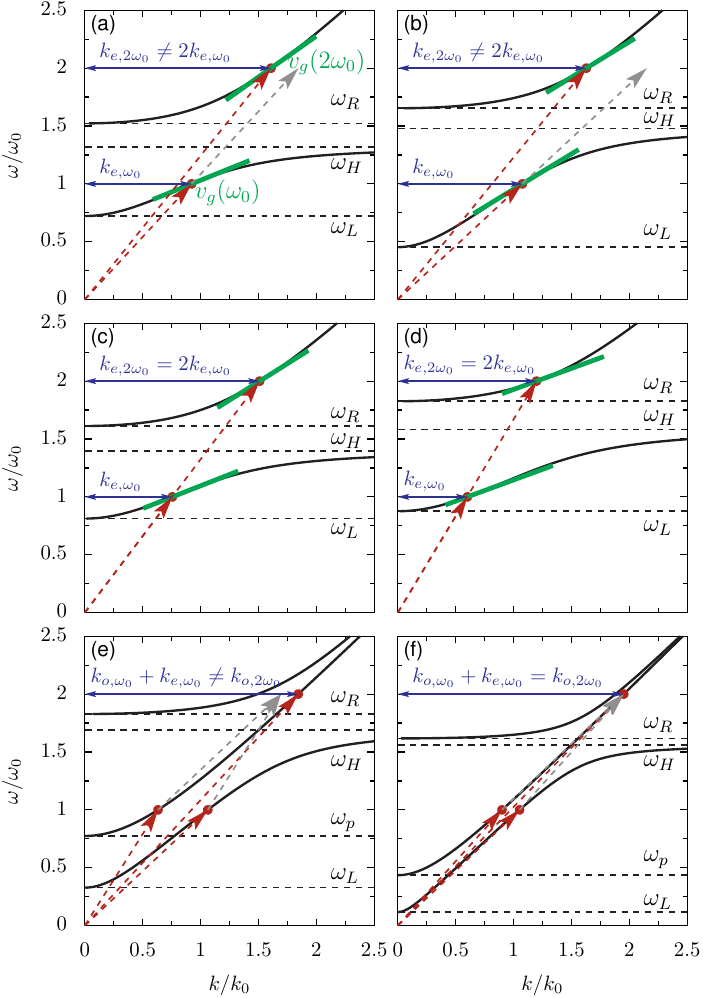}
    \caption{Dispersion diagrams for Type I SHG: 
    (a) $B_0 = 0.85$ and $N = 1.1$ where neither phase velocity nor group velocity is matched.
    (b) $B_0 = 1.2$ and $N = 0.75$ where only group velocity is matched.
    (c) $B_0 = 0.8$ and $N = 1.31$ where only phase velocity is matched.
    (d) $B_0 = 0.95$ and $N = 1.6$ where both phase velocity and group velocity of the fundamental and the second harmonic are matched.
    The green lines show the group velocity of the fundamental and the second harmonic.
    Type II SHG: (e) $B_0 = 1.5$ and $N = 0.6$ where phase-matching is not satisfied.
    (f) $B_0 = 1.5$ and $N = 0.19$ where phase-matching condition is satisfied.
    }
    \label{fig:dispersion2}
\end{figure}

\begin{figure}[t]
    \centering
    \includegraphics[width=1\linewidth]{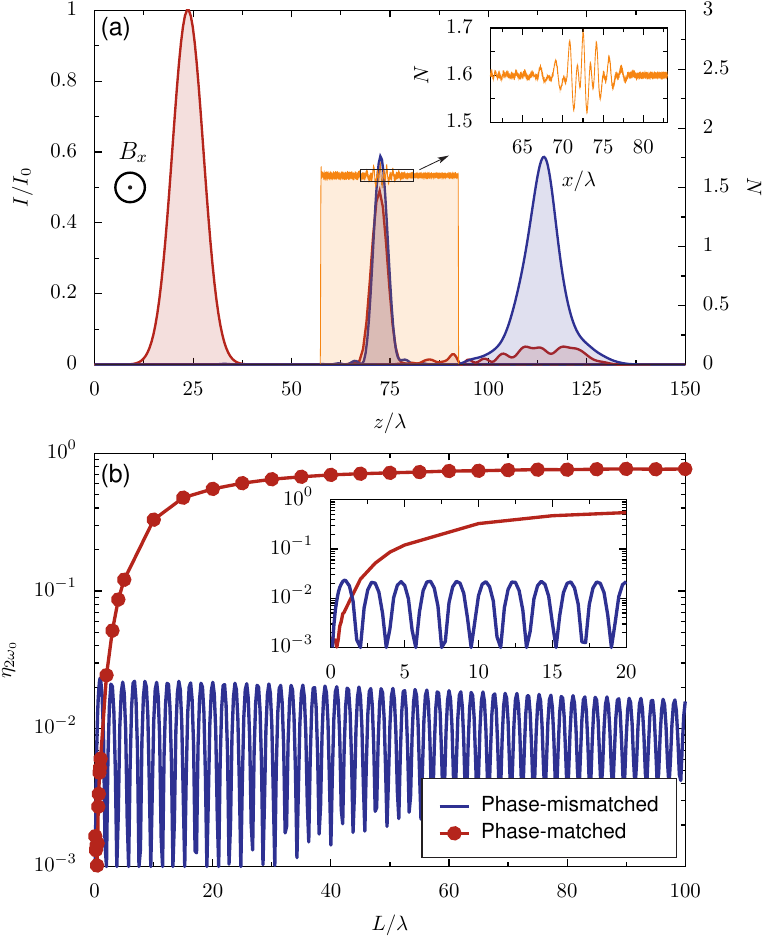}
    \caption{(a) PIC simulation of SHG in a strongly magnetized plasma with magnetic field $B_0 = 0.95$, plasma density $N = 1.6$, and plasma length $L = 35\lambda_0$. The fundamental frequency is shown in red, the second harmonic in blue, and the plasma density in orange. The evolution of the envelopes is shown as both pulses propagate through the plasma. The intensity is normalized by the initial intensity of the incident pulse. The inset shows the plasma density modulation. (b) Conversion efficiency of phase-mismatched $(B_0 = 1.2, N = 0.75)$ and phase-matched $(B_0 = 0.8, N = 1.31)$ SHG at varied plasma lengths. The inset shows the start of the efficiency oscillation in the phase-mismatched case and the efficiency growth in the phase-matched case.}
    \label{fig:PMSHG}
\end{figure}

\subsection{Type I SHG}
%
%
In Type I processes, the phase-mismatch between the fundamental and the second harmonic is $\Delta k = 2k_{e,\omega_0} - k_{e,2\omega_0}$.
To realize phase-matching, we need the refractive index of the fundamental and that of its second harmonic to be the same: $n_{e,\omega_0} = n_{e,2\omega_0}$. 
In other words, the fundamental mode $(k_{e,\omega_0}, \omega_0)$ and its second harmonic mode $(k_{e,2\omega_0},2\omega_0)$ need to be on the same line from the origin in a $\omega$-$k$ dispersion diagram.
This condition typically cannot be satisfied for a single positive definite mode.
However, with two separate branches of modes, the plasma density and magnetic field strengths can be tuned so that modes $(k_{e,\omega_0}, \omega_0)$ and $(k_{e,2\omega_0},2\omega_0)$ stay on two different branches to meet the phase-matching condition, as shown in Fig.~\ref{fig:dispersion2}cd.
For comparison, Fig.~\ref{fig:dispersion2}ab shows the case where modes $(k_{e,\omega_0}, \omega_0)$ and $(k_{e,2\omega_0},2\omega_0)$ do not have the same refractive index.
%
Setting the indices of refraction equal gives:
\begin{equation}
    \label{eq:x_wave_SHG_PMconditions}
    \sqrt{1 - \frac{\omega_p^2}{\omega_0^2}\frac{\omega_0^2-\omega_p^2}{\omega_0^2 - \omega_H^2}} = \sqrt{1 - \frac{\omega_p^2}{4\omega_0^2}\frac{4\omega_0^2-\omega_p^2}{4\omega_0^2 - \omega_H^2}}
\end{equation}

\noindent which can be rewritten, in terms of $N$ and $B_0$, as:
\begin{equation}
    \label{eq:x_wave_SHG_PMconditions_NBz}
    \sqrt{1 - N\frac{1 - N}{1 - N - B_0^2}} = \sqrt{1 -\frac{N}{4}\frac{4 - N}{4 - N - B_0^2}}
\end{equation}

\noindent We can solve for $B_0^2$ as a function of $N$:
\begin{equation}
    \label{eq:x_wave_SHG_PM_B0}
    B_0^2 = \frac{1}{N}(N-1)(4 - N)
\end{equation}

\noindent This phase-matching condition is plotted in Fig.~\ref{fig:PMconditions}a with a green line. 
The phase-matching condition requires $N > 1$ and $B_0 < 1$. 
In the limit of $N\to 1$ and $B_0\to 0$, the phase-matching condition breaks due to resonance $\omega_H\to \omega_0$, as can be seen from Eq.~\ref{eq:first_order}.

Besides phase-mismatch, which is crucial for efficient SHG, group velocity mismatch is also important, particularly for ultrashort pulses.
%
%
To achieve group velocity matching, we need $d\omega/dk$ of modes $(k_{e,\omega_0}, \omega_0)$ and $(k_{e,2\omega_0}, 2\omega_0)$ to be the same. The group velocity of the extraordinary mode can be derived as:
\begin{equation}
    \label{eq:group_velocity}
    v_g(\omega) = c\sqrt{1 - \frac{\omega_p^2}{\omega^2}\frac{\omega^2-\omega_p^2}{\omega^2 - \omega_H^2}}\left[1 - \frac{\omega_p^2(\omega_p^2 - \omega_H^2)}{(\omega^2 - \omega_H^2)^2}\right]^{-1}
\end{equation}

\noindent Setting $v_g(\omega_0) = v_g(2\omega_0)$ and $n_{\omega_0} = n_{2\omega_0}$, both phase-velocity matching and group-velocity matching are satisfied when $N \approx 1.6$ and $B_0 \approx 0.95$, as shown in Fig.~\ref{fig:dispersion2}d. 
Under this condition, we expect the fundamental to be efficiently converted to the second harmonic without much distortion in the pulse shape.
For comparison, Fig.~\ref{fig:dispersion2}b and Fig.~\ref{fig:dispersion2}c show only group velocity matching and only phase velocity matching, respectively.

We used the particle-in-cell (PIC) code EPOCH~\cite{arber2015contemporary} to simulate a laser pulse propagating through a uniform strongly magnetized cold plasma to validate the phase-matching condition.
The laser pulse had a central wavelength $\lambda_0 = 800\ \mathrm{nm}$ and a pulse duration $\tau = 35\ \mathrm{fs}$ full width at half maximum (FWHM).
The vector potential of the laser pulse was $a_0 = 0.1$, which corresponds to a peak intensity $I \approx 2.2\times 10^{16}\ \mathrm{W/cm^2} $.
%
%
The magnetic field was transverse to both the laser propagation direction and the laser polarization, and the plasma density had a rectangular profile.
We used resolutions from 150 to 600 cells/$\lambda_0$ and from 100 to 1000 particles/cell, and all the simulations are one-dimensional.

The evolution of the fundamental and the second harmonic in space and time is shown in Fig.~\ref{fig:PMSHG}a for $N = 1.6$ and $B_0 = 0.95$.
As the laser pulse propagates through the plasma, a significant fraction of its energy is converted to its second harmonic.
%
%
Fig.~\ref{fig:PMSHG}b shows how the conversion efficiency behaves as the propagation distance increases in two cases.
These two cases have the same parameters as those cases shown in Fig.~\ref{fig:eff_second_analytic}b.
The phase-mismatched case shows similar oscillatory behavior and the coherence length measured from simulations $l_c^{\mathrm{sim}}\approx 1.0\lambda_0$ agrees well with that in Fig.~\ref{fig:eff_second_analytic}b.
The phase-matched case also shows monotonic growth as the propagation distance increases, and almost $80\%$ of the energy is converted to the second harmonic.

\begin{figure}[t]
    \centering
    \includegraphics[width=1.0\linewidth]{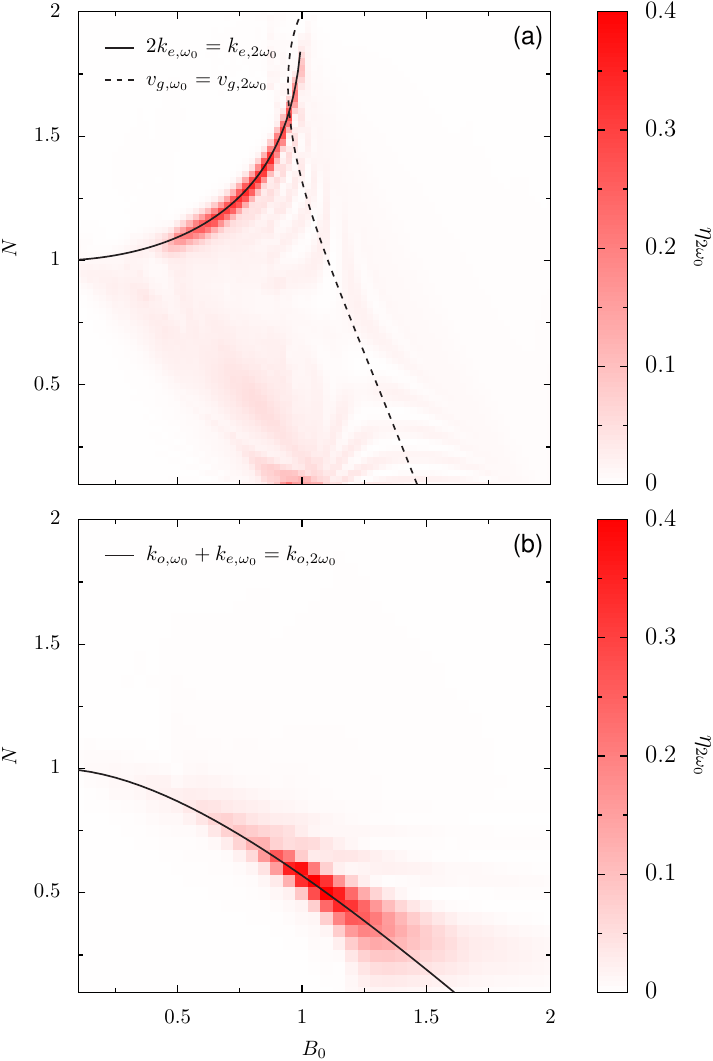}
    \caption{SHG efficiency at varied magnetic field strength $B_0$ and plasma density $N$. The plasma length is $L = 10\lambda_0$. The magnetic field is transverse to the propagation direction and (a) transverse (b) at $45^\circ$ with respect to the polarization direction of the fundamental frequency.}
    \label{fig:SHGefficiency}
\end{figure}

A scan of plasma density and magnetic field strength was carried out and the conversion efficiency of second harmonic was extracted from the energy spectra: $\eta_{2\omega_0} = T\int_{1.8\omega_0}^{2.2\omega_0}|\Tilde{E_T}(\omega)|^2d\omega/\int_{0}^\infty |\Tilde{E_T}(\omega)|^2d\omega$ where $T$ is transmittance and $|\Tilde{E_T}(\omega)|^2$ is the spectral intensity of the transmitted pulse.
Fig.~\ref{fig:SHGefficiency}a shows that the conversion efficiency has a strong peak around the parameters determined by the phase-matching conditions.
After a propagation distance of $L = 10\lambda_0$, the energy converted to the second harmonic is around $30\sim 40\%$.
This conversion efficiency scan is qualitatively similar to Fig.~\ref{fig:eff_second_analytic}a.
The agreement between the efficiency peaks in Fig.~\ref{fig:eff_second_analytic}a, Fig.~\ref{fig:SHGefficiency}a, and Eq.~\ref{eq:x_wave_SHG_PM_B0} confirms that the high conversion efficiency is a result of phase-matching.

\subsection{Type II SHG}
In Type II SHG, the magnetic field is still transverse to the light propagation direction but is not perpendicular to the polarization.
In this configuration, the fundamental frequency splits into ordinary and extraordinary modes.
Ordinary modes can scatter from the density wave induced by extraordinary mode and produce harmonic ordinary mode.
In this process, one $\omega_0$ ordinary photon and one $\omega_0$ extraordinary photon are converted to a $2\omega_0$ ordinary photon.
%
%
The nonlinear current that generates these $2\omega_0$ ordinary photons takes the form:
\begin{equation}
    \label{eq:shg_oeo_current}
    J_{x,2\omega_0} = - en_{\omega_0}v_{x,\omega_0} \propto e^{i[(k_{o,\omega_0} + k_{e,\omega_0})z - 2\omega_0t]}
\end{equation}
where $n_{\omega_0} \propto e^{i(k_{e,\omega_0}z - \omega_0 t)}$ is the density wave driven by the extraordinary mode, and the electron quiver motion driven by the ordinary mode depends on $e^{i(k_{o,\omega_0}z - \omega_0t)}$.
The second harmonic of the ordinary mode has a phase dependence $e^{i(k_{o,2\omega_0}z - 2\omega_0t)}$.
The subscript $o$ represents ordinary modes.
%
%
Therefore, to achieve phase-matched Type II second-harmonic generation, we need the current given by Eq.~\ref{eq:shg_oeo_current} to be in phase with the second harmonic: $k_{o,\omega_0} + k_{e,\omega_0} = k_{o,2\omega_0}$. 
This condition is shown in the dispersion diagram in Fig.~\ref{fig:dispersion2}f.
Mode $(k_{o,2\omega_0},2\omega_0)$ is the vector sum of modes $(k_{e,\omega_0},\omega_0)$ and $(k_{o,\omega_0},\omega_0)$.
As a comparison, Fig.~\ref{fig:dispersion2}e shows the case where the phase-matching condition is not satisfied.
In terms of refractive indices, the condition is $n_{o,\omega_0} + n_{e,\omega_0} = 2n_{o,2\omega_0}$, which can be rewritten in terms of $N$ and $B_0$ as:
\begin{equation}
    \label{eq:oeo_second}
    \sqrt{1 - N} + \sqrt{1 - N\frac{1 - N}{1 - N - B_0^2}} = 2\sqrt{1 - \frac{N}{4}}
\end{equation}

\noindent or solving for $B_0^2$ as a function of $N$:
\begin{equation}
    \label{eq:oeo_second_B0}
    B_0^2 = (1 - N)\left[1 - \frac{1}{2}\frac{N}{N + \sqrt{(4 - N)(1 - N)} - 2}\right]
\end{equation}

\noindent This phase-matching condition is plotted in Fig.~\ref{fig:PMconditions}b with a green line.
Fig.~\ref{fig:SHGefficiency}b shows PIC simulations conducted to verify the Type II phase-matching condition.
In this case, $|E_1|$ has a strength of $a_0 = 0.14$ and is polarized at $45^\circ$ with respect to the magnetic field direction.
The other simulation parameters are the same as those used for the Type I SHG simulations.
Fig.~\ref{fig:SHGefficiency}b shows that the efficiency peak is predicted well by the phase-matching condition.

\section{third harmonic generation}
The method of utilizing two branches of modes to meet the phase-matching condition can be extended to higher order harmonics.
In this section, we consider phase-matching conditions for third harmonic generation, where there are four distinct mechanisms.
%
%
Type I processes either directly convert three X-mode photons on the lower branch at frequency $\omega_0$ to one X-mode photon on the upper branch at frequency $3\omega_0$ or convert one X-mode photon on the lower branch at frequency $\omega_0$ and one X-mode photon on the upper branch at frequency $2\omega_0$ to one X-mode photon on the upper branch at $3\omega_0$.
The process $\omega_0+\omega_0+\omega_0\to 3\omega_0$ requires overdense plasma while the $\omega_0+2\omega_0\to 3\omega_0$ process requires underdense plasma.
Type II processes either convert two X-mode photons on the lower branch and one O-mode photon all at frequency $\omega_0$ to one O-mode photon at frequency $3\omega_0$, or convert one X-mode photon at frequency $\omega_0$ and one O-mode photon at frequency $2\omega_0$ to one O-mode photon at frequency $3\omega_0$.
Both Type II phase-matching conditions require underdense plasma.

%
%

\begin{figure}[t]
    \centering
    \includegraphics[width=1.0\linewidth]{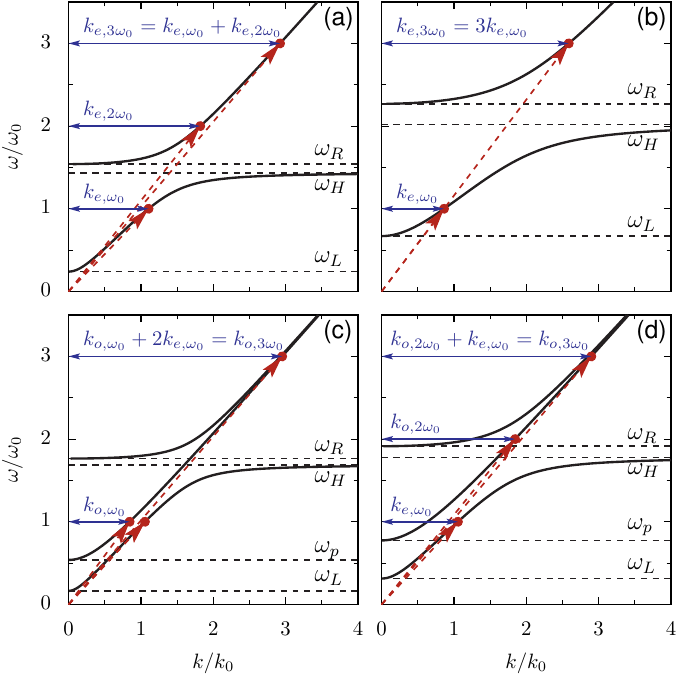}
    \caption{Dispersion diagrams for phase-matched THG of Type I: (a) $\omega_0 + 2\omega_0 = 3\omega_0$ with $B_0 = 1.3$ and $N = 0.37$. (b) $\omega_0 + \omega_0 + \omega_0 = 3\omega_0$ with $B_0 = 1.6$ and $N = 1.52$. Type II: (c) $\omega_0 + \omega_0 + \omega_0$ with $N = 0.29$ and $B_0 = 1.6$. (d) $2\omega_0 + \omega_0 = 3\omega_0$ with $N = 0.6$ and $B_0 = 1.6$.}
    \label{fig:dispersion3}
\end{figure}
\subsection{Type I THG}
%
%
We have already demonstrated that as the pulse propagates through the plasma, $2\omega_0$ photons will be generated.
Just as an X-mode with frequency $\omega_0$ induces a density wave with frequency $\omega_0$, its second harmonic induces a density wave with frequency $2\omega_0$.
The density wave induced by the second harmonic will have a wavenumber $ k_{e,2\omega_0}$.
%
%
%
Therefore, the nonlinear current that contributes to THG takes the form:
\begin{equation}
\label{eq:THG_current}
\begin{split}
    J_{y,3\omega_0} &= -e(n_{\omega_0}v_{y,2\omega_0} + n_{2\omega_0}v_{y,\omega_0})
\end{split}
\end{equation}
where the first term $n_{\omega_0}v_{y,2\omega_0}\propto e^{i((k_{e,2\omega_0} + k_{e,\omega_0})z - 3\omega_0t)}$ and the second $n_{2\omega_0}v_{y,\omega_0}\propto e^{i(3k_{e,\omega_0}z - 3\omega_0t)}$. The two terms in the nonlinear current produce two distinct resonance conditions for phase-matched THG.
The first is $3k_{e,\omega_0} = k_{e,3\omega_0}$, where three $\omega_0$ photons are converted to one $3\omega_0$ photon.
This phase-matching condition is shown in Fig.~\ref{fig:dispersion3}b.
%
The phase-matching condition in terms of $N$ and $B_0$ is:
\begin{equation}
    \label{eq:omega+omega+omega=3omega}
    \sqrt{1 - N\frac{1 - N}{1 - N - B_0^2}} = \sqrt{1 - \frac{N}{9}\frac{9 - N}{9 - N - B_0^2}}
\end{equation}
\noindent which simplifies to:
\begin{equation}
    \label{eq:omega+omega+omega=3omega_B0}
    B_0^2 = \frac{1}{N}(N - 1)(9 - N)
\end{equation}

\noindent This phase-matching condition requires $1<N<3$ and $0<B_0<2$.
When $N > 3$, both $\omega_0$ and $3\omega_0$ modes fall into the band-gap since $\omega_0<\omega_L$ and $\omega_H<3\omega_0<\omega_R$.
As a result, the fundamental does not propagate in the plasma.

The second phase-matching condition is $k_{e,\omega_0} + k_{e,2\omega_0} = k_{e,3\omega_0}$, which applies to the process where one $\omega_0$ photon and one $2\omega_0$ photon are converted to one $3\omega_0$ photon.
This phase-matching condition is shown in Fig.~\ref{fig:dispersion3}a.
Mode $(k_{e,3\omega_0},3\omega_0)$ is the vector sum of modes $(k_{e,\omega_0},\omega_0)$ and $(k_{e,2\omega_0},2\omega_0)$.
The phase-matching condition in terms of $N$ and $B_0$ is:
    \begin{equation}
\begin{split}
    \label{eq:omega+2omega=3omega}
    \sqrt{1 - N\frac{1 - N}{1 - N - B_0^2}} + 2\sqrt{1 - \frac{N}{4}\frac{4 - N}{4 - N - B_0^2}} \\
    = 3 \sqrt{1 - \frac{N}{9}\frac{9 - N}{9 - N - B_0^2}}
\end{split}
\end{equation}

\noindent Unlike the first condition, this phase-matching condition requires underdense plasma $0<N<1$.
Although a simple analytic expression is not available, we can numerically solve for $B_0$ for any specific $N$.
These phase-matching conditions are plotted in Fig.~\ref{fig:PMconditions}a with blue lines.

\begin{figure}[t]
    \centering
    \includegraphics[width=1.0\linewidth]{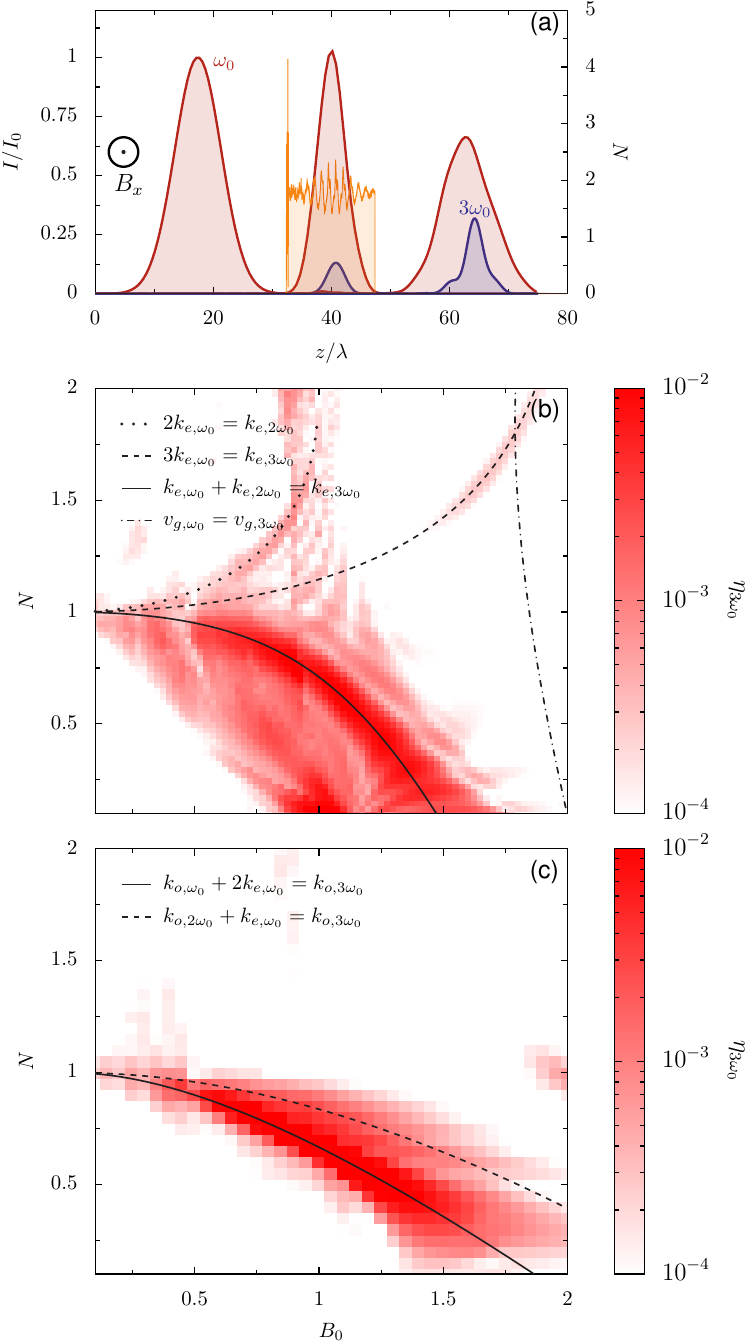}
    \caption{(a) PIC simulation of THG in a strongly magnetized plasma with $B_0 = 1.79$, $N = 1.78$, and $L = 15\lambda_0$. The fundamental frequency is shown in red, the third harmonic in blue, and the plasma density in orange. The laser pulse has a vector potential $a_0 = 0.5$ and the intensity shown is normalized by the initial intensity of the incident pulse $I\approx 5.4\times 10^{17}\ \mathrm{W/cm^2}$.
    (b)(c) THG efficiency for varied $B_0$ and $N$ with $L = 10\lambda_0$. The magnetic field is transverse to the propagation direction and (b) transverse and (c) at $45^\circ$ with respect to the polarization direction of the fundamental frequency.}
    \label{fig:THGefficiency}
\end{figure}

%
%
%
To check this analytic prediction, we filtered the energy spectra to extract conversion efficiency of third harmonic generation from the same PIC simulations presented in Fig.~\ref{fig:SHGefficiency}a.
Figure~\ref{fig:THGefficiency}b shows that the highest conversion efficiency corresponds to the phase-matching condition $k_{e,\omega_0} + k_{e,2\omega_0} = k_{e,3\omega_0}$, and we find a maximum conversion efficiency of around $1\%$ for a propagation distance of $L = 10\lambda_0$.
This resonance condition corresponds to a two stage process.
In the first stage, two $\omega_0$ photons are converted to a $2\omega_0$ photon.
In the second stage, one $\omega_0$ photon and one $2\omega_0$ photon are converted to a $3\omega_0$ photon.
Under the phase-matching condition, the second stage of this process is phase-matched, but the first stage is not.
This suggests that this third harmonic generation may be limited by the number of $2\omega_0$ photons.
Despite this, the conversion efficiency is still orders of magnitude higher than that under other parameters.
There is also a local efficiency peak under the phase-matching condition $3k_{e,\omega_0} = k_{e,3\omega_0}$, and a peak along the phase-matching condition of Type I SHG.
%
%
%
%
This last peak occurs because many more $2\omega_0$ photons are generated in the first stage when SHG is phase-matched, allowing more $3\omega_0$ photons to be produced, even though the second stage is not phase-matched.

Higher conversion efficiency can be achieved with stronger laser pulses.
An incident pulse with a peak intensity $I\approx 5.4\times 10^{17}\ \mathrm{W/cm^2}$ is shown in Fig.~\ref{fig:THGefficiency}a.
The density and magnetic field are chosen such that both phase-matching condition $3k_{e,\omega_0} = k_{e,3\omega_0}$ and group velocity matching condition $v_{g}(\omega_0) = v_g(3\omega_0)$ are satisfied.
After a propagation distance of $L = 15\lambda_0$, around $14\%$ of the incident energy is converted to the third harmonic.

\subsection{Type II THG}
In this section, we discuss phase-matching conditions for Type II THG.
%
%
We consider the nonlinear current that contributes to THG of ordinary photons.
The current comes from the coupling between the density wave and the electron quiver motion, which takes the form:
\begin{equation}
    \label{eq:THG_current_oeo}
    \begin{split}
        J_{x,3\omega_0} &= -e(n_{\omega_0}v_{x,2\omega_0} + n_{2\omega_0}v_{x,\omega_0}) 
    \end{split}
\end{equation}
%
where the first term $n_{\omega_0}v_{x,2\omega_0}\propto e^{i((k_{o,2\omega_0}+k_{e,\omega_0})z - 3\omega_0t)}$ and the second $n_{2\omega_0}v_{x,\omega_0}\propto e^{i((k_{o,\omega_0}+2k_{e,\omega_0})z - 3\omega_0t)}$. There are also two sets of phase-matching conditions.
One is $k_{o,\omega_0} + 2k_{e,\omega_0} = k_{o,3\omega_0}$, where one ordinary $\omega_0$ photon and two extraordinary $\omega_0$ photons are converted to one ordinary $3\omega_0$ photon.
Figure~\ref{fig:dispersion3}c shows the dispersion diagram for this process.
%
%
The mode $(k_{o,3\omega_0},3\omega_0)$ is the vector sum of modes $(k_{o,\omega_0},\omega_0)$ and $(2k_{e,\omega_0}, 2\omega_0)$.
All three branches of modes are involved.
The phase-matching condition in terms of $N$ and $B_0$ is:
\begin{equation}
    \label{eq:oeo_third}
    \sqrt{1 - N} + 2\sqrt{1 - N\frac{1 - N}{1 - N - B_0^2}} = 3\sqrt{1 - \frac{N}{9}}
\end{equation}
This expression can be simplified to:
\begin{equation}
    \label{eq:oeo_third_B0}
    B_0^2 = (1 - N)\left[1 - \frac{2N}{N + \sqrt{(9-N)(1-N)} - 3} \right]
\end{equation}

The other phase-matching condition is $k_{o,2\omega_0} + k_{e,\omega_0} = k_{o,3\omega_0}$, where one ordinary $2\omega_0$ photon and one extraordinary $\omega_0$ photon are converted to one ordinary $3\omega_0$ photon.
In this case, mode $(k_{o,3\omega_0},3\omega_0)$ is the vector sum of modes $(k_{o,2\omega_0},2\omega_0)$ and $(k_{e,\omega_0}, \omega_0)$, as shown in Fig.~\ref{fig:dispersion3}d.
The phase-matching condition is:
\begin{equation}
    \label{eq:oeo_third_2omega+omega}
    2\sqrt{1 - \frac{N}{4}} + \sqrt{1 - N\frac{1 - N}{1 - N - B_0^2}} = 3\sqrt{1 - \frac{N}{9}}
\end{equation}
A simplified expression is:
\begin{equation}
    \label{eq:oeo_third_2omega+omega_B0}
    B_0^2 = (1-N)\left[1 - \frac{1}{2}\frac{N}{N + \sqrt{(9-N)(4-N)} - 6}\right]
\end{equation}

Figure~\ref{fig:PMconditions}b shows these two phase-matching conditions in $N$-$B_0$ space with blue lines.
Figure~\ref{fig:THGefficiency}c shows that the global efficiency peak corresponds to the phase-matching condition $k_{o,\omega_0} + 2k_{e,\omega_0} = k_{o,3\omega_0}$. 
%
%
%
The other phase-matched process $k_{o,2\omega_0} + k_{e,\omega_0} = k_{o,3\omega_0}$ involves $2\omega_0$ ordinary photons and $\omega_0$ extraordinary photons.
The process where $2\omega_0$ ordinary photons are generated is not phase-matched.
%
%
Despite this, the efficiency is still orders of magnitude higher than other regions where neither phase-matching condition is satisfied.

\begin{table*}[t]
\caption{Phase-matching conditions for different harmonic generation processes}
\label{tbl:params}
\begin{ruledtabular}
\begin{tabular}{l c c c c}
\noalign{\smallskip}
&{\bf Type I}\footnote{Type I involves only the extraordinary mode. The subscript $e$ denotes the extraordinary mode.} &{\bf Phase-matching condition}\footnote{The function $f_n(N,B_0)$ is defined as: $f_n(N,B_0) = \sqrt{1 - \frac{N}{n^2}\frac{n^2 - N}{n^2 - N - B_0}}$.}& {\bf Applicable range}\\
\noalign{\smallskip}
\hline
\noalign{\medskip}
{\bf SHG}& $2k_{e,\omega_0}= k_{e,2\omega_0}$ & $B_0^2 = \frac{1}{N}(N-1)(4-N)$ & $N\in (1,2),\ B_0\in(0,1)$\footnote{Both group velocity matching and phase-matching are satisfied with $N \approx 1.6$ and $B_0 \approx 0.95$.}\\
\noalign{\medskip}
\hline
\noalign{\medskip}
{\bf THG}& $3k_{e,\omega_0} = k_{e,3\omega_0}$ &$ B_0^2 = \frac{1}{N}(N - 1)(9 - N)$& $N\in (1,3),\ B_0\in (0,2)$ \footnote{Both group velocity matching and phase-matching are satisfied with $N \approx 1.78$ and $B_0 \approx 1.79$.} \\
\noalign{\medskip}
&$k_{e,2\omega_0} + k_{e,\omega_0} = k_{e,3\omega_0}$ &$f_1 + 2f_2 = 3f_3$& $N\in(0,1)$\\
\noalign{\medskip}
\hline
\noalign{\medskip}
{\bf FHG}& $4k_{e,\omega_0} = k_{e,4\omega_0}$&$B_0^2 = \frac{1}{N}(N - 1)(16 - N)$& $N\in (1,4),\ B_0\in (0,3)$ \\
\noalign{\medskip}
&$k_{e,2\omega_0} + 2k_{e,\omega_0} = k_{e,4\omega_0}$ & $f_1 + f_2 = 2f_{4}$ & $N\in(0,1)$\\
\noalign{\medskip}
&$k_{e,3\omega_0} + k_{e,\omega_0} = k_{e,4\omega_0}$ &$f_1 + 3f_3 = 4f_{4}$& $N\in(0,1)$\\
\noalign{\medskip}
\hline
\noalign{\medskip}
&{\bf Type II}\footnote{Type II involves both the extraordinary mode and the ordinary mode. The subscript $o$ denotes the ordinary mode.} &{\bf Phase-matching condition}&{\bf Applicable range}\\
\noalign{\smallskip}
\hline
\noalign{\medskip}
{\bf SHG}& $k_{o,\omega_0} + k_{e,\omega_0}= k_{o,2\omega_0}$ & $B_0^2 = (1 - N)\left[1 - \frac{1}{2}\frac{N}{N + \sqrt{(4 - N)(1 - N)} - 2}\right]$ & $N\in (0,1),\ B_0\in (0,\sqrt{3})$ \\
\noalign{\medskip}
\hline
\noalign{\medskip}
{\bf THG}& $k_{o,\omega_0} + 2k_{e,\omega_0} = k_{o,3\omega_0}$ &$B_0^2 = (1 - N)\left[1 - \frac{2N}{N + \sqrt{(9-N)(1-N)} - 3} \right]$& $N\in (0,1),\ B_0\in (0,2)$ \\
\noalign{\medskip}
&$k_{o,2\omega_0} + k_{e,\omega_0} = k_{o,3\omega_0}$ &$B_0^2 = (1-N)\left[1 - \frac{1}{2}\frac{N}{N + \sqrt{(9-N)(4-N)} - 6}\right]$& $N\in (0,1),\ B_0\in (0,\sqrt{7})$\\
\noalign{\medskip}
\hline
\noalign{\medskip}
{\bf FHG}& $k_{o,\omega_0} + 3k_{e,\omega_0}= k_{o,4\omega_0}$ &$B_0^2 = (1-N)\left[1 - \frac{1}{2}\frac{9N}{N + \sqrt{(16-N)(1-N)}-4}\right]$& $N\in (0,1),\ B_0\in (0,\sqrt{5})$ \\
\noalign{\medskip}
&$k_{o,2\omega_0} + 2k_{e,\omega_0} = k_{o,4\omega_0}$ & $B_0^2  = (1-N)\left[1 - \frac{2N}{N + \sqrt{(16-N)(4-N)} - 8}\right]$ & $N\in (0,1),\ B_0\in (0,3)$ \\
\noalign{\medskip}
&$k_{o,3\omega_0} + k_{e,\omega_0} = k_{o,4\omega_0}$ &$B_0^2  = (1-N)\left[1 - \frac{1}{2}\frac{N}{N + \sqrt{(16-N)(9-N)} -12}\right]$& $N\in (0,1),\ B_0\in (0,\sqrt{13})$ \\
\end{tabular}
\end{ruledtabular}
\end{table*}

\begin{figure}[t]
    \centering
    \includegraphics[width=1.0\linewidth]{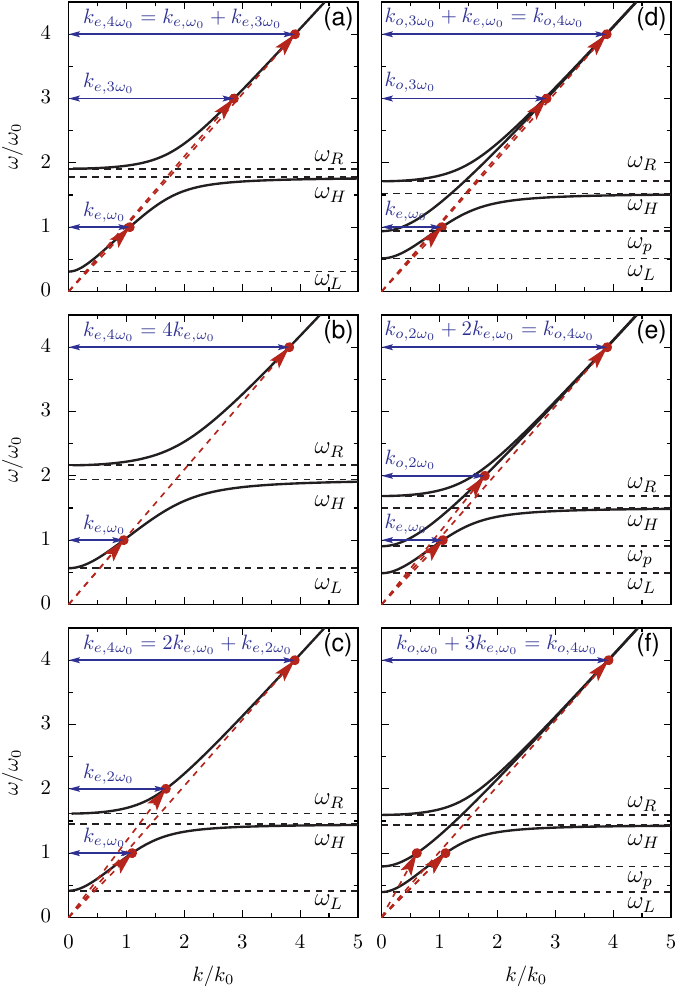}
    \caption{Dispersion diagrams for phase-matched FHG.
    (a)-(c) Type I with (a) $N = 0.59$ and $B_0 = 1.6$, (b) $N = 1.22$ and $B_0 = 1.6$, and (c) $N = 0.67$ and $B_0 = 1.2$.
    (d)-(f) Type II with (d) $N = 0.88$ and $B_0 = 1.2$, (e) $N = 0.82$ and $B_0 = 1.2$, and (f) $N = 0.63$ and $B_0 = 1.2$.
    }
    \label{fig:dispersion4}
\end{figure}

\section{fourth harmonic generation}
We can further extend this analysis to find phase-matching conditions for fourth harmonic generation.
The nonlinear current consists of more terms for FHG, which results in six possible pathways for the frequency up-conversion.
%
%
Three phase-matched Type I processes are identified: (1) One X-mode photon on the lower branch at frequency $\omega_0$ and one X-mode photon on the upper branch at frequency $3\omega_0$ are converted to a X-mode photon on the upper branch at frequency $4\omega_0$; (2) Two X-mode photons on the lower branch at frequency $\omega_0$ and one X-mode photon on the upper branch at $2\omega_0$ are converted to one X-mode photon on the upper branch at $4\omega_0$; (3) Four X-mode photons on the lower branch at $\omega_0$ are converted to one X-mode photon on the upper branch at $4\omega_0$.
For Type II, there are also three phase-matched processes: (1) One X-mode photon on the lower branch at $\omega_0$ and one O-mode photon at $3\omega_0$ are converted to one O-mode photon at $4\omega_0$; (2) Two X-mode photons on the lower branch at $\omega_0$ and one O-mode at $2\omega_0$ are converted to one O-mode photon at $4\omega_0$; (3) Three X-mode photons on the lower branch at $\omega_0$ and one O-mode at $\omega_0$ are converted to one O-mode photon at $4\omega_0$.

%

\subsection{Type I FHG}
%
%
%
Both $2\omega_0$ and $3\omega_0$ photons are generated by a propagating X-mode pulse, so density waves oscillating with frequency $2\omega_0$ and $3\omega_0$ exist.
The nonlinear current that contributes to fourth harmonic generation then takes the form:
\begin{equation}
    \label{eq:FHG_current}
    \begin{split}
        J_{y,4\omega_0} &= -e(n_{\omega_0}v_{y,3\omega_0} + n_{2\omega_0}v_{y,2\omega_0} + n_{3\omega_0}v_{y,\omega_0})\\
    \end{split}
\end{equation}
%
where the first term $n_{\omega_0}v_{y,3\omega_0}\propto e^{i((k_{e,\omega_0} + k_{e,3\omega_0})z - 4\omega_0t)}$, the second $n_{2\omega_0}v_{y,2\omega_0}\propto e^{i(( 2k_{e,\omega_0} + k_{e,2\omega_0})z - 4\omega_0t)}$, and the third $n_{3\omega_0}v_{y,\omega_0}\propto e^{i(4k_{e,\omega_0}z - 4\omega_0t)}$. The three terms give three phase-matching conditions.
We first consider $4k_{e,\omega_0} = k_{e,4\omega_0}$.
Fig.~\ref{fig:dispersion4}b shows the corresponding dispersion diagram.
Four $\omega_0$ photons are converted to a $4\omega_0$ photon, which requires:
%
\begin{equation}
    \label{eq:omega+omega+omega+omega=4omega_B0}
    B_0^2 = \frac{1}{N}(N - 1)(16 - N)
\end{equation}

The corresponding dispersion diagrams for the other two processes are shown in Fig.~\ref{fig:dispersion4}ac.
The phase-matching conditions can be derived by considering $k_{e,2\omega_0} + 2k_{e,\omega_0} = k_{e,4\omega_0}$ and $k_{e,3\omega_0} + k_{e,\omega_0} = k_{e,4\omega_0}$.
%
%
%
%
The phase-matching condition for $k_{e,2\omega_0} + 2k_{e,\omega_0} = k_{e,4\omega_0}$ expressed in $N$ and $B_0$ is:
 \begin{equation}
    \begin{split}
        \label{eq:omega+omega+2omega=4omega}
    \sqrt{1 - N\frac{1 - N}{1 - N - B_0^2}} + \sqrt{1 - \frac{N}{4}\frac{4 - N}{4 - N - B_0^2}}\\ = 2\sqrt{1 - \frac{N}{16}\frac{16 - N}{16 - N - B_0^2}}
    \end{split}
\end{equation}
and for $k_{e,3\omega_0} + k_{e,\omega_0} = k_{e,4\omega_0}$ is:
\begin{equation}
\begin{split}
    \label{eq:omega+3omega=4omega}
    \sqrt{1 - N\frac{1 - N}{1 - N - B_0^2}} + 3\sqrt{1 - \frac{N}{9}\frac{9 - N}{9 - N - B_0^2}}\\ = 4\sqrt{1 - \frac{N}{16}\frac{16 - N}{16 - N - B_0^2}}
\end{split}
\end{equation}

\noindent The purple lines in Fig.~\ref{fig:PMconditions}a show these phase-matching conditions in $N$-$B_0$ space.

\begin{figure}[t]
    \centering
    \includegraphics[width=1.0\linewidth]{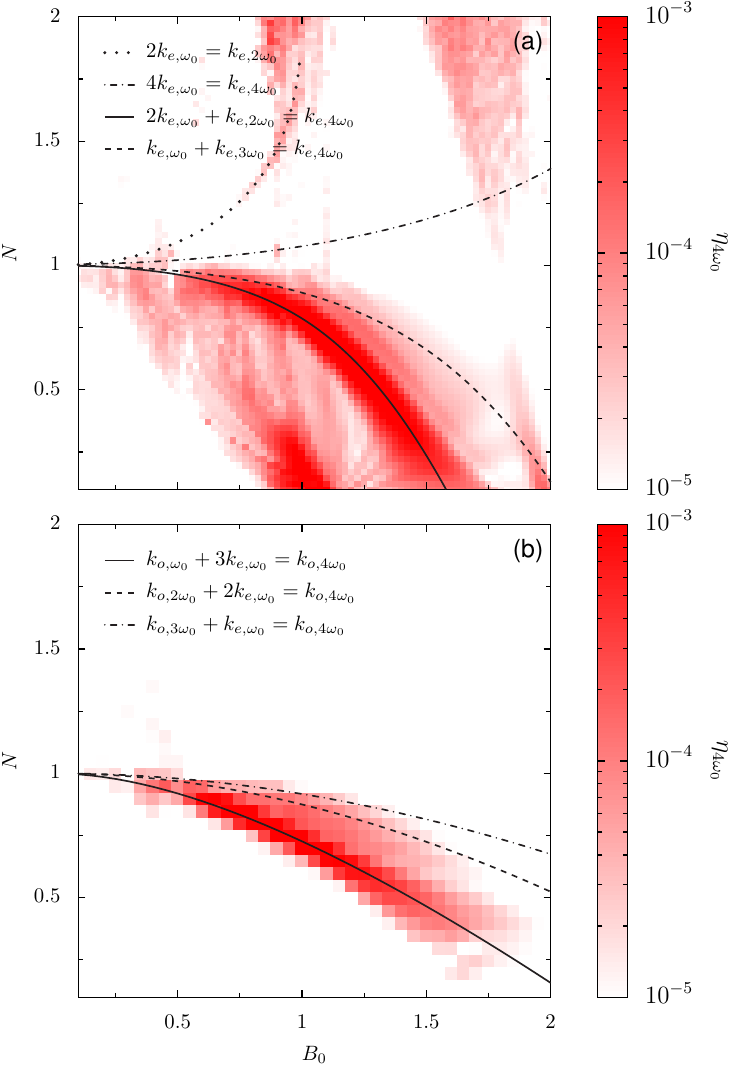}
    \caption{FHG efficiency for varied $B_0$ and $N$. The plasma length is $L = 10\lambda_0$. The magnetic field is transverse to the propagation direction and (a) transverse (b) at $45^\circ$ with respect to the polarization direction of the fundamental frequency.}
    \label{fig:FHGefficiency}
\end{figure}

Figure~\ref{fig:FHGefficiency}a shows the efficiencies of FHG calculated with PIC simulations at different $N$ and $B_0$.
%
%
The efficiency under the phase-matching condition given by Eq.~\ref{eq:omega+omega+2omega=4omega} is orders of magnitude higher than that under other parameters, and about $0.1\%$ of the incident energy is converted to the fourth harmonic.
The efficiency under the phase-matching conditions given by Eq.~\ref{eq:omega+3omega=4omega} and Eq.~\ref{eq:x_wave_SHG_PM_B0} also shows a local peak, where Eq.~\ref{eq:x_wave_SHG_PM_B0} is the phase-matching condition of Type I SHG.
Fourth harmonic generation is easier when more $2\omega_0$ photons are available.
The phase-matching condition $4k_{e,\omega_0} = k_{e,4\omega_0}$ does not produce a significant signature in these PIC simulations.
Similar to third harmonic generation, we can achieve higher conversion efficiency of fourth harmonic generation by using laser pulses with higher intensities. 
More than 2\% of the incident energy can be converted to its fourth harmonic for an incident pulse with a peak intensity $I \approx 2\times 10^{17}\ \mathrm{W/cm^2}$, after a propagation distance of $L = 15\lambda_0$.
The plasma density and magnetic field strength are $N = 0.5$ and $B_0 = 1.31$, where the phase-matching condition $2k_{e,\omega_0} + k_{e,2\omega_0} = k_{e,4\omega_0}$ is satisfied.

\begin{figure}[t]
    \centering
    \includegraphics[width=1.0\linewidth]{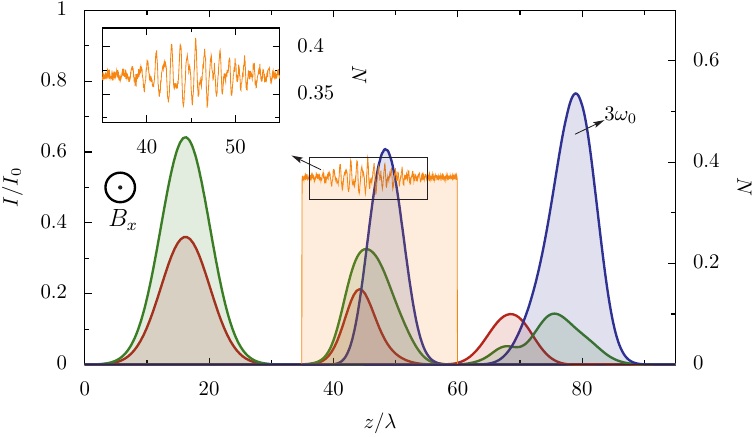}
    \caption{PIC simulation of THG in a strongly magnetized plasma with $B_0 = 1.3$, $N = 0.37$, and $L = 25\lambda_0$. The two colors have vector potentials $a_{\omega_0} = 0.06$ and $a_{2\omega_0} = 0.04$ respectively. The intensities are normalized by the intensity corresponding to $\omega = \omega_0$ and $a_0 = 0.1$, which has the same energy as the input two-color beam. The envelope of fundamental frequency is shown in red, the second harmonic in green, the third harmonic in blue, and the plasma density in orange.}
    \label{fig:twocolorschematic}
\end{figure}

\begin{figure}[t]
    \centering
    \includegraphics[width=1.0\linewidth]{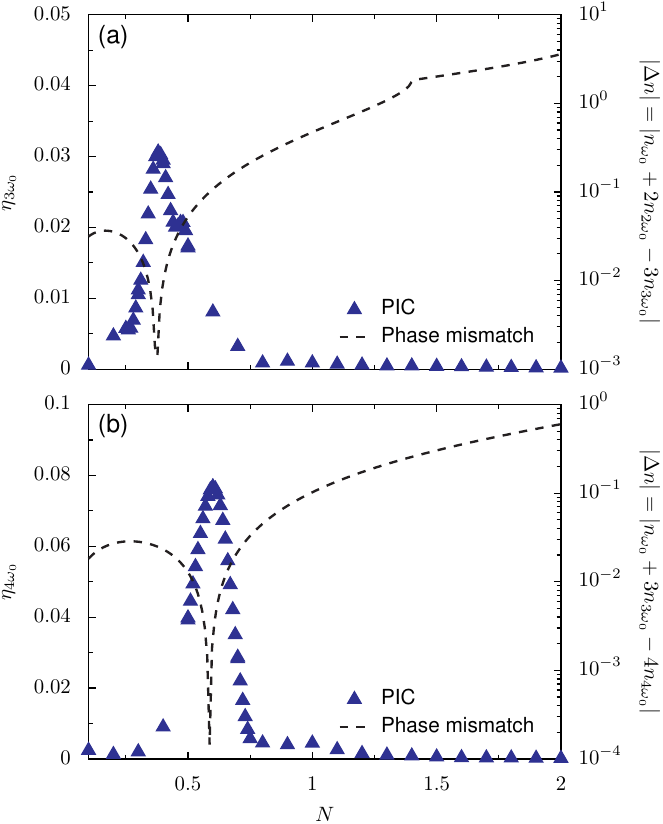}
    \caption{Conversion efficiency and phase-mismatch of Type I THG and FHG with two-color beams at varied plasma densities. (a) THG with $a_0 = 0.1$, $a_{2\omega_0} = 0.01$, and $B_0 = 1.3$. (b) FHG with $a_0 = 0.1$, $a_{3\omega_0} = 0.01$ and $B_0 = 1.6$. The plasma length is $L = 25\lambda_0$.}
    \label{fig:TwoColorEfficiency}
\end{figure}

\subsection{Type II FHG}
The nonlinear current that drives FHG with Type II phase-matching takes the form:
\begin{equation}
    \label{FHG_current_oeo}
    \begin{split}
        J_{x,4\omega_0} &= -e(n_{\omega_0}v_{x,3\omega_0} + n_{2\omega_0}v_{x,2\omega_0} + n_{3\omega_0}v_{x,\omega_0})\\
    \end{split}
\end{equation}
%
where the first term $n_{\omega_0}v_{x,3\omega_0}\propto e^{i((k_{e,\omega_0} + k_{o,3\omega_0})z - 4\omega_0t)}$, the second $ n_{2\omega_0}v_{x,2\omega_0}\propto e^{i((2k_{e,\omega_0} + k_{o,2\omega_0})z - 4\omega_0t)}$, and the third $n_{3\omega_0}v_{x,\omega_0}\propto e^{i((3k_{e,\omega_0} + k_{o,\omega_0})z - 4\omega_0t)}$. This current suggests that three phase-matching conditions exist, including $3k_{e,\omega_0} + k_{o,\omega_0} = k_{o,4\omega_0}$, $2k_{e,\omega_0} + k_{o,2\omega_0} = k_{o,4\omega_0}$, and $k_{e,\omega_0} + k_{o,3\omega_0} = k_{o,4\omega_0}$.
The dispersion diagrams corresponding to each of these processes are shown in Fig.~\ref{fig:dispersion4}def.
Following the same method as before, we can derive the phase-matching conditions for each of these processes.
For $3k_{e,\omega_0} + k_{o,\omega_0} = k_{o,4\omega_0}$:
\begin{equation}
    \label{eq:FHG_oeo_omega+omega+omega_omega}
    B_0^2 = (1-N)\left[1 - \frac{1}{2}\frac{9N}{N + \sqrt{(16-N)(1-N)}-4}\right]
\end{equation}
For $2k_{e,\omega_0} + k_{o,2\omega_0} = k_{o,4\omega_0}$:
\begin{equation}
    B_0^2  = (1-N)\left[1 - \frac{2N}{N + \sqrt{(16-N)(4-N)} - 8}\right]
     \label{eq:FHG_oeo_2omega+omega+omega}
\end{equation}
and for $k_{e,\omega_0} + k_{o,3\omega_0} = k_{o,4\omega_0}$:
\begin{equation}
    B_0^2  = (1-N)\left[1 - \frac{1}{2}\frac{N}{N + \sqrt{(16-N)(9-N)} -12}\right]
     \label{eq:FHG_oeo_3omega+omega}
\end{equation}

%
\noindent These phase-matching conditions, which all require underdense plasma, are shown in Fig.~\ref{fig:PMconditions}b with purple lines.
Figure~\ref{fig:FHGefficiency}b shows a conversion efficiency scan for Type II FHG in $N$-$B_0$ space, considering again the same PIC simulations presented in Fig.~\ref{fig:SHGefficiency}b, filtered now for the fourth harmonic.
The maximum appears around the phase-matching condition given by Eq.~\ref{eq:FHG_oeo_omega+omega+omega_omega}, and about $0.1\%$ of the incident energy is converted to the fourth harmonic.
%
%
Although it is hard to see signatures corresponding to the other two phase-matching conditions here, these mechanisms show up mpre clearly when driven by two-color beams, as shown in the next section.

The agreement demonstrates that phase-mismatch in harmonic generation can be minimized in strongly magnetized plasmas provided that the interaction parameters are carefully selected.
Table~\ref{tbl:params} summarizes all the phase-matching conditions presented in this work.

 \section{Two-color beams}
When a single-color pulse propagates through the plasma, third and fourth harmonic generation processes are limited by the number of $2\omega_0$ and $3\omega_0$ photons.
If a two-color pulse that consists of $\omega_0$ and $2\omega_0$ or $\omega_0$ and $3\omega_0$ is sent through the plasma, the conversion efficiency of third or fourth harmonic under the phase-matching conditions is expected to be substantially higher. 
Therefore, we check the performance of phase-matched processes driven by two-color beams.
Figure~\ref{fig:twocolorschematic} shows a PIC simulation of a two-color beam propagating through a plasma with $64\%$ of incident energy in its second harmonic.
The two colors are both polarized transverse to the magnetic field so only extraordinary waves are excited.
The phase-matching condition involved here is $k_{e,\omega_0} + k_{e,2\omega_0} = k_{e,3\omega_0}$.
%
%
Around 70\% of the incident energy is converted to third harmonic after a propagation distance of $L = 25\lambda_0$.
%

\begin{figure}[t]
    \centering
    \includegraphics[width=1.0\linewidth]{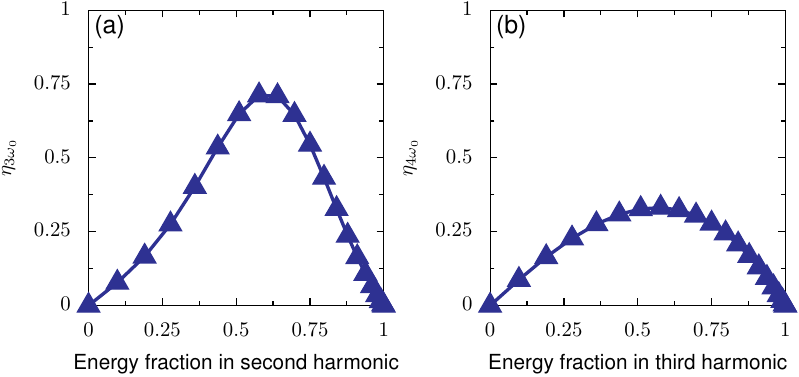}
    \caption{Conversion efficiency of type 1 THG and FHG using two-color beams with varied energy fraction in the second color. (a) THG with $N = 0.37$ and $B_0 = 1.3$. (b) FHG with $N = 0.59$ and $B_0 = 1.6$. The plasma length is $L = 25\lambda_0$. $a_0 = 0.1$ if there is no second color.}
    \label{fig:TwoColorEnergyFraction}
\end{figure}

%
To show that this efficient harmonic generation is a result of phase-matching, we simulate two-color pulses propagating through a plasma with varied plasma densities.
We first consider Type I third- and fourth-harmonic generation processes.
Two-color beams consisting of $\omega_0$ and $2\omega_0$ for THG and $\omega_0$ and $3\omega_0$ for FHG were used.
Both colors were polarized perpendicularly to the magnetic field.
For THG, the strength of the $2\omega_0$ color was $a_{2\omega_0} = 0.01$, where $a_{2\omega_0} = eE_2/m_e(2\omega_0)c$.
For FHG, the strength of the $3\omega_0$ color was $a_{3\omega_0} = 0.01$, where $a_{3\omega_0} = eE_3/m_e(3\omega_0)c$.
$E_2$ and $E_3$ are the maximum electric fields of the second and third harmonic.
As shown in Fig.~\ref{fig:TwoColorEfficiency}, the efficiency maxima line up with the minima of the phase-mismatch for both THG and FHG.
With less than $10\%$ energy in the second color, the conversion efficiency of THG and FHG with a two-color beam is orders of magnitude higher than that with a single-color beam.
Fig.~\ref{fig:TwoColorEnergyFraction} shows that the conversion efficiency peaks when the energy fraction in the second color is around $60\%$.
The maximum conversion efficiencies of third and fourth harmonic are more than $70\%$ and $30\%$, respectively.

\begin{figure}[t]
    \centering
    \includegraphics[width=1.0\linewidth]{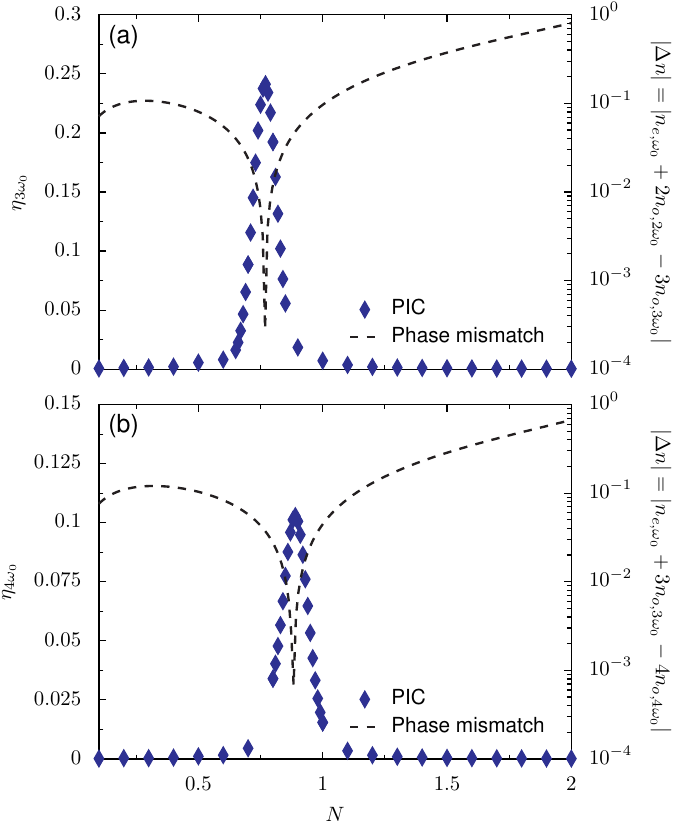}
    \caption{Conversion efficiency and phase-mismatch of Type II THG and FHG with two-color beams at varied plasma densities. (a) THG with $a_0 = 0.1$, $a_{2\omega_0} = 0.05$ and $B_0 = 1.2$. (b) FHG with $a_0 = 0.1$, $a_{3\omega_0} = 0.033$ and $B_0 = 1.2$. The plasma length is $L = 25\lambda_0$.}
    \label{fig:TwoColorEfficiency_oeo}
\end{figure}

We next consider Type II harmonic generation with two-color beams.
In this case, the fundamental frequency polarization is perpendicular to the magnetic field and the second color polarization is parallel to the magnetic field.
As the plasma density varies, a good agreement between the maximum of the conversion efficiency and the minimum of the phase-mismatch can be seen in Fig.~\ref{fig:TwoColorEfficiency_oeo}.
The efficiency of Type II harmonic generation with a two-color beam is also orders of magnitude higher than that with a single-color beam.

In addition to using two-color beams for efficient FHG, we can cascade two phase-matched SHG processes.
%
%
In the first stage, a single-color pulse with the central frequency $\omega_0$ is sent through the plasma and a second harmonic beam with the central frequency $2\omega_0$ comes out. 
The plasma length can be adjusted so that most energy is converted to the second harmonic. 
The second harmonic beam is then sent through the second frequency up-conversion stage, with four times the plasma density, and twice the magnetic field strength of the first stage.
With two cascaded phase-matched SHG steps, the fundamental pulse can be up-shifted to its fourth harmonic.
In the best case, the overall conversion efficiency may be the square of a single SHG step.



\section*{Discussion}


%
%

%

%
%
%

Our analysis of phase-matching conditions can be extended to fifth, sixth, and higher order harmonics.
For higher order harmonic generation, we expect more phase-matched processes of both types to be possible.
However, the conversion efficiency with a single color beam would potentially decrease significantly.
%
%
Based on our results, this issue may be improved with two-color beams.
%
%
For fifth harmonic generation, one may consider using a two-color beam to generate third harmonic first.
Then using another two-color beam containing second and third harmonic to realize fifth harmonic generation.
Sixth harmonic generation may be realized by cascading a two-color third harmonic generation stage and a second harmonic generation stage.
The corresponding phase-matching conditions can be derived following the same methods as those presented here.

Previous studies have suggested that strongly magnetized plasmas have potential in reproducing astrophysical environments in a laboratory setting~\cite{weichman2022progress}, amplifying laser pulses to intensities beyond what can be achieved by chirped pulse amplification~\cite{jia2017kinetic,edwards2019laser,shi2019amplification}, compressing short wavelength lasers~\cite{shi2017laser}, improving direct laser acceleration of electrons~\cite{gong2020direct,arefiev2020energy}, enhancing energy gains in inertial confinement fusion experiments~\cite{chang2011fusion,slutz2012high,gotchev2009laser}, accelerating photons to high energies~\cite{stark2016enhanced}, and manipulating polarization of high power lasers.~\cite{weng2017extreme,zheng2019simultaneous}.
%
%
These applications have motivated efforts to produce strong magnetic fields $(10\ \mathrm{MG}\sim 10\ \mathrm{GG})$ in a laboratory setting, including driving strong current in coils using lasers~\cite{daido1986generation}, compressing coil type targets or driving microtube implosions~\cite{yoneda2012strong,knauer2010compressing,korneev2015gigagauss,fujioka2013kilotesla,law2016direct,murakami2020generation,shokov2021laser,santos2015laser,zhang2018generation}, driving high-intensity laser pulses through solids~\cite{wagner2004laboratory} or plasmas~\cite{lamavc2023generation}, and transferring angular momentum to plasmas using structured light~\cite{longman2021kilo}.
The actual plasma density and magnetic field strength required for phase-matched harmonic generation processes scale with the frequency, $N\propto\omega
_0^{-2}$ and $B_0\propto \omega_0^{-1}$.
The phase-matching conditions approximately require $N\approx 1$ and $B_0\approx 1$.
For high power lasers with $\lambda_0\approx 1\ \mathrm{\mu m}$, the plasma density is around $10^{21}\ \mathrm{cm^{-3}}$ and the magnetic field is around $10^4\ \mathrm{T}$.
%
%
For $\mathrm{CO_2}$ lasers with $\lambda_0 = 10.6\ \mathrm{\mu m}$, the plasma density is around $10^{19}\ \mathrm{cm^{-3}}$, and the magnetic field required is around $10^3\ \mathrm{T}$ ($10\  \mathrm{MG}$).
Although these field strengths are high, it is possible that near-future experiments could approach these values, making diagnostics for these field strengths valuable.

%
%

%
%
%
%
%

In comparison to relativistic high harmonic generation from solid targets~\cite{gibbon1996harmonic,lichters1996short,edwards2020x}, the other mechanism for harmonic generation using plasmas, the processes presented in this work allow precisely controlled production of specific harmonic order by tuning the plasma density and the magnetic field strength and is more efficient at producing narrow band radiation.

%
%
%
%
%
%

%

%
%

In conclusion, we have shown that phase-matched harmonic generation can be achieved in strongly magnetized uniform plasmas.
We have identified two types of phase-matched harmonic generation processes, distinguished by the angle between the magnetic field and the input polarization of the fundamental frequency.
The phase-matching conditions of different second, third, and fourth harmonic generation processes of both types are derived and validated using PIC simulations.
%
%
We have shown that more than $70\%$, $14\%$, and $2\%$ energy can be converted to the second, the third, and the fourth harmonic under the phase-matching condition with a single color pulse.
With two color beams, the conversion efficiency of third and fourth harmonic generation can be more than $70\%$ and $30\%$. 
In both cases, this substantial conversion efficiency results from minimizing the phase-mismatch.
Our results demonstrate that phase-matching makes harmonic generation extremely efficient in strongly magnetized plasmas, extending nonlinear optics to this regime of extreme plasma physics.

\noindent {\bf Funding.} National Science Foundation (PHY-2308641); National Nuclear Security Administration (DE-NA0004130).

\noindent {\bf Acknowledgments. }This work was partially supported by NSF Grant PHY-2308641 and NNSA Grant DE-NA0004130. The PIC code EPOCH~\cite{arber2015contemporary} is funded by the UK EPSRC grants EP/G054950/1, EP/G056803/1, EP/G055165/1 and EP/ M022463/1. The computing for this project was performed on the (Stanford) Sherlock cluster. We would like to thank Stanford University and the Stanford Research Computing Center for providing computational resources and support that contributed to these research results.

\noindent {\bf Disclosures.} The authors declare no conflicts of interest.

\noindent {\bf Data availability. }Data underlying the results presented in this paper are not publicly available at this time but may be obtained from the authors upon reasonable request.

\bibliography{reference}

\end{document}